\documentclass{edp-jp4}
\usepackage{graphicx}
\usepackage{amsmath}
\usepackage{amsfonts}
\usepackage{amssymb}
\begin{document}

\title{How to reach the collisional regime on a magnetically guided atomic beam}
\author{J. M. Vogels, T. Lahaye, C. Roos, J. Dalibard and D. Gu\'{e}ry-Odelin}
\maketitle
\begin{abstract}
In this paper, we report our progress towards the realization of a
continuous guided atomic beam in the degenerate regime. So far, we
have coupled into a magnetic guide a flux of  a few $10^{8}$
atoms/s at 60 cm/s with a propagation in the guide over more than
2 meters. At this stage, the collision rate is not high enough to
start an efficient forced evaporative cooling. Here we describe a
new approach to reach the collisional regime. It is based on a
pulsed feeding of the magnetic guide at a high repetition rate.
The overlap of the packets of atoms occurs in the guide and leads
to a continuous guided beam. We discuss different ways to increase
the collision rate of this beam while keeping the phase space
density constant by shaping the external potential.
\end{abstract}

%\author{J. M. Vogels, T. Lahaye, C. Roos, J. Dalibard and D. Gu\'ery-Odelin}
%

\address{Laboratoire Kastler Brossel, Ecole normale sup\'erieure,
24, Rue Lhomond, F-75231 Paris Cedex 05, France}

\section{Introduction}

A spectacular challenge in the field of Bose-Einstein condensation
is the achievement of a continuous beam operating in the quantum
degenerate regime. This would be the matter wave equivalent of a
CW monochromatic laser and it would allow for unprecedented
performance in terms of focalization or collimation. In
\cite{ScienceK02}, a continuous source of Bose-Einstein condensed
atoms was obtained by periodically replenishing a condensate held
in an optical dipole trap with new condensates. This kind of
technique raises the possibility of realizing a continuous atom
laser. An alternative way to achieve this goal has been proposed
and studied theoretically in \cite{Mandonnet00}. A non-degenerate,
but already slow and cold beam of particles, is injected into a
magnetic guide where transverse evaporation takes place. If the
elastic collision rate is large enough, an efficient evaporative
cooling leads to quantum degeneracy at the exit of the guide. This
scheme transposes in the space domain what is usually done in
time, so that all operations leading to the condensation are
performed in parallel, with the prospect of obtaining a much
larger output flux. In the present paper, we report our progress
along those lines, and outline our strategy to reach the required
collisional regime in the magnetic guide.

\section{Experimental requirements}

The condition for reaching degeneracy with the latter scheme can
be formulated by means of three parameters: the length $\ell$ of
the magnetic guide on which evaporative cooling is performed, the
collision rate $\gamma$ at the beginning of the evaporation stage,
and the mean velocity $v_{\mathrm{b}}$ of the beam of atoms.
Following the analysis given in \cite{Mandonnet00}, one obtains
\begin{equation}
N_{\mathrm{c}}\equiv\frac{\gamma\ell}{v_{\mathrm{b}}}>500\;.\label{cond1}%
\end{equation}
If the collision rate $\gamma$ is constant over the cooling
process, which is approximately the case for realistic conditions,
this means that each remaining atom at the end of the guide has
undergone $N_{\mathrm{c}}$ elastic collisions during its
collisional propagation through the magnetic guide.

Some conclusions can already been drawn from the inequality
(\ref{cond1}). One needs to operate in a long magnetic guide, at
very low mean velocity, and the collision rate should be as high
as possible at the beginning of the evaporation. The criterion
(\ref{cond1}) can be recast in terms of the temperature $T$, the
incoming flux $\phi$, and the strength
$\lambda$ of the linear transverse confining potential: $N_{\mathrm{c}}%
\sim\phi\lambda^{2}v_{\mathrm{b}}^{-2}T^{-3/2}$. We consequently need to start
with a large incoming flux at low velocity and at very low temperature.

\section{Continuous injection}

\begin{center}
\begin{figure}[ptb]
\includegraphics[width=11cm]{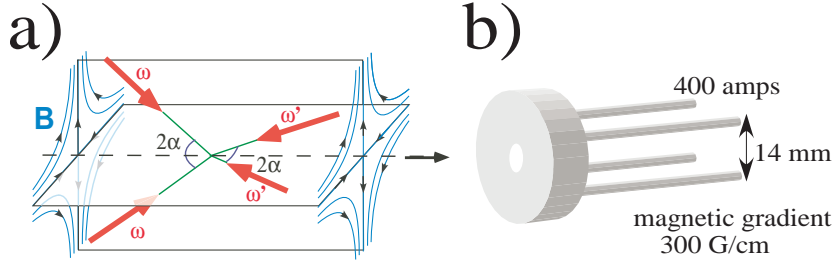}\caption{1a) Laser configuration
of the injector magneto-optical trap. 1b) Entrance of the magnetic
guide. The cylindrical hollow metal piece allows for the
connection of the electrical currents and  cooling water
circulating into the four copper
tubes.}%
\end{figure}
\end{center}

\subsection{Injector}

In our experiment, we produce a slow atomic beam with
magneto-optical trap called the {\it injector} that has been
described in detail elsewhere \cite{cren}. It is based on four
laser beams in a tetrahedral configuration, superimposed with a
two-dimensional magnetic quadrupole field (see figure 1a). The
transverse gradient is typically 10 G/cm, and each beam has a
power of 25 mW and a waist of 15 mm. This geometry is dictated by
the requirement of a free axis (no confinement) needed for the
propagation of the beam. This kind of trap provides trapping in
two dimensions and cooling in three dimensions. The absence of
trapping along one direction makes this trap very sensitive to
local imbalance of intensity, and one has to use intensity
stabilized and spatially filtered beams. The frequency of one pair
of beams is adjusted with respect to the frequency of the other
pair in such a way so as to perform the cooling in a
longitudinally moving frame with an adjustable velocity
$v_{\mathrm{b}}$ ranging from 0 to 3 m/s. This technique is
reminiscent of the one used in atomic fountain clocks to launch
the atoms through the cavity.

\subsection{How does one load atoms in the injector?}

To load the injector magneto-optical trap we have investigated two
methods.

 First, we have loaded the atoms from a low-pressure
background gas. By controlling the temperature of the rubidium
reservoir and the size of its aperture, we could vary the
$^{87}$Rb pressure from $10^{-9}$ to a few $10^{-8}$ mbar. This
method suffers from the following drawback. In order to increase
the flux, one would like to increase the vapor pressure. However,
this also increases the losses due to collisions between atoms
coming out of the injector MOT with thermal background atoms. The
best results we obtained, subject to this trade-off, were a flux
of $10^{9}$ atoms/s at 2 m/s for a $^{87}$Rb  vapor pressure of
$P_{87}=10^{-8}$ mbar \cite{cren}.

In order to avoid the losses from the background pressure, we have
implemented a new setup relying on two magneto-optical traps in
two different vacuum chambers connected by a differential vacuum
tube. In the first chamber, we used a two-dimensional trap in the
presence of a relatively high rubidium pressure ($10^{-7}$ mbar)
\cite{walraven98,pfau02}. It generated a beam of pre-cooled atoms
with a flux of typically $10^{10}$ atoms/s with an average
velocity of 40 m/s \cite{roos}. This beam was captured by the
injector located in the second chamber with background pressure of
the order of $10^{-9}$ mbar. At this level the residual pressure
did not affect any more the output flux. For technical reasons,
the pre-cooled beam was along the axis of the injector
magneto-optical trap for which no gradient of magnetic field
existed. Actually, this geometry was not very efficient for good
capturing of the incoming atoms. Indeed, atoms are decelerated
more efficiently in the presence of a magnetic gradient, since the
latter enables full on-resonant deceleration while the atoms are
being captured. As a result the loading efficiency of the injector
is very anisotropic. For an incoming flux of $10^{10}$ atoms/s
(along the trap axis), we estimate to recapture a few $10^8$
atoms/s despite the large intensity used for the laser beams.

In order to deal with this anisotropic loading rate, we are
currently implementing a Zeeman slower with a recirculating oven.
This should provide a high flux ($>10^{10}$ atoms/s) which arrives
perpendicular to the longitudinal axis of the injector to take
advantage of the transverse gradient of the trap in order to more
efficiently capture atoms.

\subsection{Continuous injection of atoms into the magnetic guide}

\begin{center}
\begin{figure}[ptb]
\includegraphics[width=11cm]{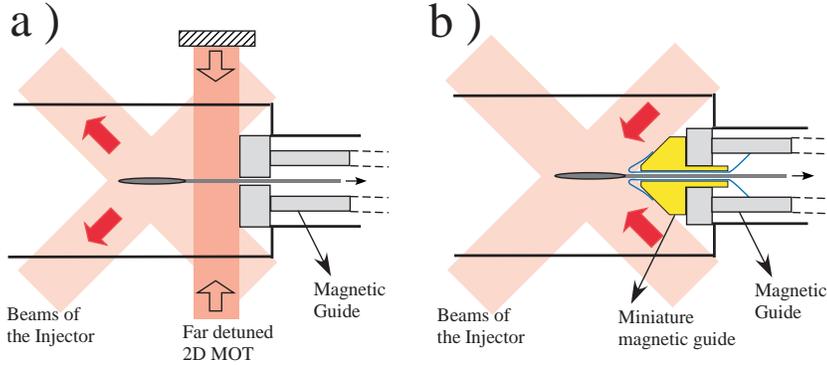}\caption{2a) Use of a two-dimensional
far-detuned magneto-optical trap to transfer the atoms from the
injector to the magnetic guide. Alternatively, we use a miniature
magnetic guide 2b) placed at the entrance of the magnetic guide to
facilitate the
transfer.}%
\end{figure}
\end{center}

Different magnetic guides have already been investigated experimentally for
guiding ultracold atoms
\cite{Schmiedmayer95,Denschlag99,Goepfert99,Key00,Dekker00,Teo01,Sauer01,Hinds99}%
. Since our application requires a long guiding time, we developed
a 2.3 meter long magnetic guide. It consisted of four copper tubes
($\O_{\mathrm{ext}}=$ 6 mm and $\O_{\mathrm{int}}=$ 4 mm) in
quadrupolar configuration joined at the beginning of the guide by
a hollow metal cylinder (see fig. 1b) which allowed for the
circulation of high current and cooling water from tube to tube.
The distance between the tubes at the entrance was 14 mm. For this
separation between the tubes, we obtained a gradient of 300 G/cm
with a current of 400 A. The magnetic field produced by the guide
fell off sufficiently fast and did not affect the injector
performance. However, there existed an intermediate region from
the output of the injector to the entrance of the magnetic guide
where no significant confinement was provided. In this free flight
region, the jet of atoms expanded outwards in space and became
dilute. As a consequence, the collision rate decreased
dramatically. To circumvent this problem, we investigated two
different strategies to confine the atoms in this region.

\subsubsection{Far-detuned magneto-optical trap}

In the first set of experiments, the magnetic guide was 0.6 meters
long \cite{cren} and the injector was vapor-loaded. To confine the
atoms in the intermediate region we superimposed a far-detuned
($\delta\sim - 7 \Gamma$) two-dimensional magneto-optical trap
(see Fig. 2a). We operated the injector in the continuous mode and
achieved the continuous feeding of the magnetic guide with a flux
of $3\times 10^{8}$ atoms/s at 2 m/s. We evaluated the total
number of collisions $N_{c}$ undergone by an atom reaching the end
of the magnetic guide to be $0.3$. This first experiment
demonstrated the feasibility of the continuous loading with high
flux. However, it remained very far from the requirement
(\ref{cond1}) to start an efficient evaporative cooling of the
guided beam.

\subsubsection{Miniature guide}

In a second set of experiments we used a 2.3 meter long guide and
the injector was loaded from a two-dimensional magneto-optical
trap in a different vacuum chamber, as described above. The
far-detuned two-dimensional magneto-optical trap did not provide
an efficient coupling of atoms into the guide at low velocity due
to strong recoil heating in the longitudinal direction. For this
reason, we developed a miniature magnetic guide on a conical
supporting structure. The shape was chosen in order to maintain a
good optical access for the beams of the injecting magneto-optical
trap. This miniature guide \cite{roos} was added at the entrance
of the magnetic guide as depicted in Fig. 2b. With this improved
setup, we were able to couple a flux of 1.5$\times 10^{8}$ atoms/s
at a very low velocity of the order of 60 cm/s, thereby reaching a
total number of collisions per atom in the guide of $N_{c}=3$.

We emphasize that the low mean velocity $v_{\mathrm{b}}$
requirement implies the need for a fine control of the height of
the guide. For instance, a 2 mm vertical displacement of the guide
would be sufficient to stop atoms moving at 20 cm/s.

\section{Pulsed mode}

\subsection{Motivation}

In a regular Bose-Einstein condensation experiment, atoms are
first captured by a magneto-optical trap, are cooled further by
means of a molasses and are polarized before being trapped
magnetically. The shape of the magnetic trap is adjusted to the
size and temperature of the cloud to achieve good mode matching.
This method allows one to obtain the highest possible initial
phase space density and collision rate.

Strictly speaking, this mode-matching procedure cannot be applied
to a continuous beam. Indeed, the jet of atoms produced by the
injector cannot be cooled efficiently by a molasses since the
gradient of the magnetic field and the small detuning should
remain to capture atoms continuously, and furthermore atoms are
not transferred instantaneously into the magnetic guide. Another
significant problem encountered, especially at low velocities, was
the unavoidable presence of repumping light at the entrance of the
magnetic guide, which could kick atoms out of the trapped
spin-state. Repumping light rescattered by atoms did not appear to
be a significant contribution, and, if necessary, could be reduced
by providing off-resonant repumping light at higher intensities.
These problems, which arise for a continuous mode, made it
difficult to achieve a high number of collisions $N_{\mathrm{c}}$
\cite{cren}.

\subsection{Pulsed injection}

To optimize the transfer of the atoms from the injector into the
magnetic guide, we are currently investigating the possibility to
operate in pulsed mode. The atoms are first captured in the
injector set to zero launch velocity. During this stage the
detuning and transverse magnetic field gradient are optimized for
efficient capturing. Then the detuning of the injector is adjusted
to a larger value and its intensity is turned down so that the
temperature of the atoms is decreased by the molasses effect.
Also, during this stage the injector is ramped to a finite
velocity (between 0.5 m/s and 1 m/s) to launch the atoms.
Subsequently the atoms are optically pumped to the proper Zeeman
sublevel. Next, they are magnetically trapped by a two-dimensional
quadrupole field strong enough ($\sim$ 60 G/cm) to uphold them
from falling due to gravity. A longitudinal bias field (a few
gauss) is temporarily provided to ensure mode matching during this
catching stage and is then removed adiabatically, which causes a
first compression. As soon as this packet has left the capture
region of the injector and enters the long magnetic guide, another
packet is prepared, and so on. In this way we can still capture a
large fraction of the atoms, while optimally treating the packet
for injection into the waveguide. These packets eventually overlap
after a propagation into the magnetic guide of the order of 50 cm
for realistic experimental conditions, leading to a truly
continuous beam afterwards.

\section{How does one increase the collision rate of the guided beam?}

The collision rate of the beam can be modified by changing the
external potential experienced by the atoms. In our experiment,
the compression is twofold. First, it is provided by the removal
of the longitudinal bias field used for mode matching. At this
stage the confinement is changed from initially harmonic to
linear. Secondly, the atoms are compressed by entering and
propagating in a tapered magnetic guide in which the transverse
confinement increases as the atoms progress into the guide. In
practice, the latter compression is achieved by decreasing the
distance between the guide tubes.

The increase of the strength of the transverse confinement would
ideally be slow enough to ensure the validity of thermodynamical
adiabatic conditions. In this case, the phase space density and
the enthalpy remain constant through the constriction, and the
collision rate $\gamma$ as well as $N_{c}$ increase significantly.
Upon compression, the temperature $T$ and the thermal velocity
$v_{\mathrm{th}}=(k_{B}T_{0}/m)^{1/2}$ increase, while the
velocity of the beam decreases by up to a factor $2\sqrt{2}$
\cite{lahaye}.

The initial injection velocity should be chosen such that the
thermal velocity remains lower than the mean velocity during the
compression: $M\equiv v_{\mathrm{b}}/v_{\mathrm{th}}>1$. In the
hydrodynamic regime, $M$ would be the Mach number within a
numerical factor of order unity. Actually, if the compression is
too strong, with $M$ reaching unity, atoms are reflected. As a
consequence, a stronger compression requires a larger initial
ratio $M$, but nevertheless a net gain in $N_{c}$ is achieved. The
initial temperature is a crucial factor, since it determines the
initial density of the cloud, as well as how slow we can inject
the atoms into the guide: if the injection velocity is kept as low
as possible, {\it i.e.}, with $M$ fixed, $N_{c}$ scales as
$\sim\phi T^{- 5/2}$. We stress that evaporation during
compression could be beneficial, since it enhances the ratio $M$,
allowing for an even lower injection velocity. However, if the
optimal injection ratio $M$ is higher due to experimental
limitations in the injection region, we still have the
possibility to locally tilt the guide in order to increase $N_{c}%
$\cite{lahaye}.

We recently realized the first part of the pulsed injection
procedure and were able to obtain an initial collision rate of 4
s$^{-1}$ in packets magnetically trapped in the injection region
at a transverse gradient of 60 G/cm. With this performance,
$N_{c}$ could exceed $100$ in the $2.3$ m long guide, by taking
advantage of the compression, which is encouraging.

\section{Conclusion}

We have summarized our experimental results concerning the
continuous loading of a high flux of atoms into a long magnetic
guide. We have briefly discussed a new strategy based on a pulsed
feeding of the guide. To implement an efficient evaporative
cooling we need a high number of collisions per atom through their
propagation into the guide. In the above we have considered the
possibility of increasing the collision rate by adiabatically
modifying the external potential experienced by the atoms. We have
emphasized that adiabatic modifications require the velocity of
the beam to be significantly larger than the thermal velocity.

\end{document}